\documentclass[conference]{IEEEtran}

\usepackage[cmex10]{amsmath}
\numberwithin{equation}{section}
\numberwithin{figure}{section}

\title{Formal methods and software engineering
for DL\\
\huge{Security, safety and productivity for DL systems development}
}

\author{

\IEEEauthorblockN{Ga\'etan J.D.R. Hains\IEEEauthorrefmark{1} and Arvid Jakobsson}
\IEEEauthorblockA{Huawei Parallel and Distributed Algorithms Lab.\\
Huawei Paris Research Center\\
Boulogne Billancourt, France\\
  \{ gaetan.hains, arvid.jakobsson\}@huawei.com }
\IEEEauthorrefmark{1}{{\small On leave from LACL, Univ. Paris-Est Cr\'eteil, France}}

\and

\IEEEauthorblockN{Youry Khmelevsky\IEEEauthorrefmark{2}}
\IEEEauthorblockA{LIP6, Database and Machine Learning Department\\ Sorbonne University, Paris, France\\
youry.khmelevsky@lip6.fr}
\IEEEauthorrefmark{2}{{\small Extended study leave from Comp.Sc., Okanagan College, Canada }}

}

\begin{document}
\maketitle

\begin{abstract}
Deep Learning (DL) techniques are now widespread and being integrated into many important systems. Their 
classification and recognition abilities ensure their relevance for multiple application domains. 
As machine-learning that relies on training instead of algorithm 
programming, they offer a high degree of productivity. But they can be 
vulnerable to attacks and the verification of their correctness is only just emerging as a scientific and 
engineering possibility.  
This paper is a major update of a previously-published survey, attempting to cover all 
recent publications in this area. 
It also covers an even more recent trend, namely the design of domain-specific languages for 
producing and training neural nets. 
\end{abstract}

\section{Introduction}
\setlength{\fboxrule}{2pt} 
As research unit of a leading vendor of information and communication systems, 
Huawei's Central Software Institute (CSI) 
is developing high-performance deep learning (DL) systems for image classification \cite{park2004content} and 
other image recognition functions.  
In application domains like self-driving cars \cite{kisavcanin2017deep}, correct operation (safety) 
and attack resistance (security) of DL systems has become critical. 
The engineering of neural networks (NN) is less well understood than for general software: 
despite a relatively static and clean structure, their functionality depends on numerical 
parameters that are extracted from ad-hoc datasets and complex hand-made layer topologies built from 
signal-processing operators and treshold or "activation" nodes. 
As a result, a neural network's behavior depends mostly on its numerical values, and its 
use in embedded systems is not amenable to verification by control-flow. 

A ray of hope in this bleak outlook, 
recent research has found a partial substitute to full NN specification and verification in the 
form of novel stability analysis techniques. 
Such techniques ensure that a small change in input (image, sound or pattern) produces a negligible 
change in output e.g. no change in the embedded system's behavior. 
Several groups have shown how to adapt model checking for this purpose, others have designed 
special-purpose linear solvers for it, and the computational feasibility of this analysis has been 
improving. 
It remains to see whether trust in NN inference will reach the level required of safety-critical 
applications. 
But a clear trend has been set to improve the understanding and engineering of this very popular type 
of machine learning. 
 
This paper is a major update of a previously-published survey \cite{Hains2018}, attempting to cover all 
recent publications in this area. 
Research on safety of DL had produced two papers per year in the period 2003-2014. 
We then found three directly-relevant publications in 2015, seven in 2016, sixteen in 2017 and a 
relative slowdown with 9 publications in 2018. 
This peek corresponds with the arrival and proof-of-concept for feasible static verification of 
NN stability, hence their protection against so-called adversarial attacks. 
Our survey also covers a few papers on an even more recent trend, 
namely the design of programming languages for producing and training neural nets. 
The work covered constitutes in our opinion the first generation of tools for neural network software 
engineering. 
 
 The next sections survey existing work on 
 \begin{itemize}
  \item Attacks against DL systems
  \item Testing, training and monitoring DL systems for safety
  \item The verification of DL systems
\end{itemize}

Then we survey recent work and propose new work in the design of programming tools for DL. 

%% Section: SECURITY: ATTACKS AND THEIR PREVENTION
\section{SECURITY: ATTACKS AND THEIR PREVENTION}

An adversarial example for a NN classifier is a slightly perturbed input that generates a different, hence wrong, 
classification from the desired one. In recent years many have been identified and specific solutions designed for each 
one. But the general problem remains of formally guaranteeing in advance the absence of adversarial example.

Carlini et al.'s paper \cite{carlini2017ground} is motivated in this manner by safety-security (absence of accidental or 
intentional adversarial examples) and the need to verify it. They introduce the notion of a ground truth, or adversarial 
example with minimal change in input value. This is useful for two things: judging the quality of an attack by comparing 
it to the ground truth, and judging the quality of a defence by the amount it increases the distortion in the new ground 
truth.

The authors of \cite{lu2017no} present and articulate technical arguments that appear to show that intentional 
adversarial examples can be countered, in the area of image processing, by a kind of ``multi-sensor" approach. Like 
attacks on face recognition can be countered by 3D or multiple-angle 2D images, adversarial examples would become 
ineffective in the presence of multiple-angle or time-sequenced images of the same object(s).

%%%%%%%%%%%%%%%%%%
% TESTING, TRAINING  AND MONITORING FOR SAFETY 
%%%%%%%%%%%%%%%%%%
\section{TESTING, TRAINING  AND MONITORING FOR SAFETY}

Concrete progress has been made by authors who propose to adapt training and testing with specific safety-conscious 
properties and techniques. 

The survey paper by B. Taylor et al. \cite{taylor2003verification} takes a very general human-level definition of AI 
safety. It defines eight very challenging wanted properties of machine learning systems like NN but most of them relate 
to the human application of DL systems so, in our opinion, they are premature to consider before the science and 
engineering of DL becomes more mature. One of their eight properties is more amenable to purely technical developments ``
inductive ambiguity identification" with special case ``active learning". An active learner can interact with humans 
during its learning phase so as to ask them for additional data (e.g. images) that would break some automatically 
detected ambiguity in classification. Active learning can thus be considered a design goal for improving the safety of 
DL systems.

The authors of \cite{tian2018deeptest} consider the application of an (unrelated) automatic testing tool called DeepTest 
to self-driving cars. It can be considered an elementary but meaningful tool for structured testing. As such it has the 
advantages and limitations of testing methods: easy to design and implement, incomplete by design. 

Leofante, Pulina and Tacchella \cite{leofante2016learning} present recent work in the definition and verification of 
machine-learning safety, namely the guarantee that the input-output function defined by a trained system will behave ``
according to specification". They also quote model-checking results for verifying this property, its computational costs 
but do not detail the methodology for doing this. Their notion of global correctness is based on stability: limited 
input sample variations lead to limited output variations. This is a well-defined and apparently verifiable type of 
specification, but it does open two related and deep questions: how can designers be certain that their reference 
datasets are in some sense correct and complete? How to choose the metric that measures the input or output variations? 
The notion of active learning, presented in \cite{taylor2003verification} could lead to a practical solution to the 
first question. But the general problem of global correctness certainly needs more powerful mathematical tools than 
stability theory: NNs must interact with general algorithms, if only for such operations as sorting results, and the 
whole system's correct and complete specification is thus a classical pre-condition, post-condition pair of local 
expressions on the system state. In the (very common) application area of image processing NN-specific predicates could 
specify that image recognition is, for example rotation invariant. To the best of our knowledge this problem of mixing 
signal-processing with software specification is unexplored. Stability predicates would then be an important but 
incomplete tool to ensure system correctness.

 Wicker, Huang and Kwiatkowska \cite{wicker2018feature} present a sophisticated approach that allows black-box testing 
of NNs i.e. with consideration of features being detected but ignorance of the NN's structure. They search a game space 
where an agent adversary attempts to use normally/fool/randomly use the detection of features. The method is considered 
competitive with white-box methods. 

Yerramalla, Mladenovski and Fuller \cite{yerramalla2005stability} applies continuous control theory to design a monitor 
for ensuring that ``unstable" learning can be detected. Their notion of stability is specific to an application where a 
fixed dataset of images is replaced by an airplane's onboard NN that is trained dynamically through in-flight cameras. 
This work can be considered as mathematical support for dynamically generated datasets, or abstractly: dynamically 
generated specifications for the DL system. 

%%% NEW TEXT in Red Box
%\begin{tcolorbox}[colback=red!2!white, colframe=red!75!black]{ %Red Box

Wu et al. \cite{wu2018game}  proposed a two-player turn-based game framework for the verification of deep neural 
networks with provable guarantees, and to evaluate pointwise robustness of neural networks in safety-critical 
applications such as traffic sign recognition in self-driving cars. They  developed a software tool DeepGame, and 
demonstrated its applicability on networks and dataset benchmarks.\\

Gehr et al. \cite{dd2018ai2} present AI$^2$ a scalable analyzer for deep neural networks, a system able to certify 
convolutional and large fully connected networks. Based on over approximation, AI$^2$ can automatically prove safety 
properties (e.g., robustness) of realistic neural networks (e.g., convolutional neural networks) with an extensive 
evaluation on 20 neural networks.\\

Black and Ribeiro \cite{black2016sate} developed the Ockham Sound Analysis Criteria to recognize static analyzers whose 
findings are always correct.  In Static Analysis Tool Exposition (SATE) V, only one tool was reviewed.\\

Georgakis et al. \cite{georgakis2017synthesizing} investigated the ability of using synthetically generated composite 
images for training state-of-the-art object detectors, especially for object instance detection.
They superimpose 2D images of textured object models into images of real environments at variety of locations and 
scales. They demonstrate the effectiveness of these object detector training strategies on two publicly available 
datasets, the GMUKitchens \cite{georgakis2016multiview} and the Washington RGB-D Scenes v2 \cite{lai2014unsupervised}.\\

Hinterstoisser, Lepetit and Wohlhart \cite{hinterstoisser2017pre} demonstrated how to  train effectively modern object 
detectors with synthetic images only. They ``freeze'' the layers responsible for feature extraction to generic layers 
pre-trained on real images, and train only the remaining layers with plain OpenGL rendering. They have shown that by 
freezing a pre-trained feature extractor they are able to train state-of-the-art object detectors on synthetic data 
only, and freezing the feature extractor gives a huge performance boost.\\

Jang, Wu and Jha \cite{jang2017objective} focused on attacks by adversarial perturbation. They present a simple 
gradient-descent based algorithm for finding adversarial perturbations, which performs well in comparison to existing 
algorithms. 
They present a novel metric based on computer-vision algorithms for quantifying the difference between an image and its 
perturbation.
%} \end{tcolorbox} %End of the Red Box

%%% NEW TEXT in Red Box
%\begin{tcolorbox}[colback=red!2!white, colframe=red!75!black]{ %Red Box

Leofante et al. \cite{leofante2018automated} propose an automated reasoning technique and a comprehensive categorization 
of existing approaches for the automated verification of neural networks. In their opinion the automated verification of 
NNs could be the new driving force for theoretical and practical advancements in Automated Reasoning and, at the same 
time, ML could benefit from powerful verification techniques to generate proofs of correctness for NNs.
%} \end{tcolorbox} %End of the Red Box

But again, testing is by design an incomplete approach and the ``specification" of a DL system relies on the 
experimental definition of its training dataset. 
              
%%%%%%%%%%%%%%%%%%
% VERIFICATION AND SIMULATION
%%%%%%%%%%%%%%%%%%
\section{VERIFICATION AND SIMULATION}

Other authors have investigated formal and even automatic methods for safety verification. This line of research has 
been accelerating in recent years. 

Broderick \cite{broderick2005adaptive} uses simulation in the area of flight on-board online-learning NNs. It does not 
take a formal approach to verification but applies statistical techniques. The white paper \cite{van2017challenges} 
defines high-level requirements for ``formal" (mathematically-based) verification of similar systems from the point of 
view of control theory. 

Fuller, Yerramalla and Cukic \cite{fuller2006stability} model the learning of a NN as a dynamical system where training 
adjustments are discrete differential equations on the states that are neurons and weights. Lyapunov stability analysis 
is then applicable to detect stable states in the dynamical system. Stability in this theory thus amounts to the absence 
of adversarial examples. It is shown how to apply this concept to (shallow) NNs of fixed topology and also to dynamic 
ones. 

Survey paper \cite{bunel2017piecewise} compares methods for verifying NNs with piecewise linear structures. It compares 
methods based on SMT solvers, mixed integer programming and a new branch-and-bound method. The tools are able to verify 
100-500 properties for networks for 2-6 layers. Correctness is defined as a form of stability and verification, in 
theory exhaustive testing, is accelerated by assuming piecewise-linear state spaces. 

Katz et al. \cite{gopinath2017deepsafe,katz2017reluplex} treat Rectified Linear Units-based (ReLU) NN systems.
The NN system and a domain specific safety specification is modelled as an SMT formula. The system is verified
using a version of the simplex algorithm modified to handle the non-linearities introduced by the ReLU-functions. 
However, their use-case has a
well-defined safety specification, which is not the case in other domains such as image recognition. Furthermore, 
scalability is a concern for this technique.

Cheng, N\"uhrenberg and Ruess \cite{cheng2017verification} verify DNNs by translating non-linear (input-output) 
constraints generated by ReLU activation functions using big-M encoding. Then standard techniques for linear 
optimization are applied to verification.

In \cite{ehlers2017formal}, an optimization technique is proposed to accelerate verification problems that are difficult 
for SMT and ILP solvers. It assumes so-called feed-forward NNs that allow the addition of a global linear approximation 
of the overall network behavior. 

Blog entry \cite{goodfellow2017challenge} is a general discussion of the importance of safety for DL systems, with 
arguments in favour of formal verification as opposed to testing. 

Huang et al. \cite{huang2017safety} present SMT-based work on verifying the absence of adversarial inputs in Feed-
forward multi-layer neural networks. The paper contains many convincing examples of such perturbed images. The 
verification method finds adversarial inputs, if they exist, for a given region and a family of manipulations.

 Katz et al. published in \cite{katz2017towards} their efforts to prove adversarial robustness of NNs. They propose a 
new notion of "global robustness" quantifying the robustness of a DNN.  Intuitively, a network is globally robust if 
any two neighbours in the input are also neighbors in the output. Robustness is thus a non-limit form of continuity as 
in: \[
   d_1(x,y)\leq\delta\longrightarrow d_2(NN(x),NN(y))\leq\varepsilon
% d_1(x,y)\leqslant\delta\longrightarrow d_2(NN(x),NN(y))\leqslant\varepsilon
 \]

where $NN$ is the neural net's inference function, $d_1$ is a standard metric in the input domain, $d_2$ a suitable 
metric in the output domain and $\delta, \varepsilon$ are experimentally chosen error bounds where $\varepsilon$ could 
be zero, e.g. if the output is a discrete space of features. They then show how to encode this property and verify it 
using Reluplex. However, it is challenging to verify, and the result only extends to DNN with a few dozen nodes.

Narodytask et al. \cite{narodytska2017verifying} present the first exact Boolean representation of a deep NN so that a 
binarized network is faithfully represented as a Boolean formula. They are then able to leverage the high efficiency of 
modern SAT solvers for the formal and automatic verification of the NNs behavior, in particular resistance to 
adversarial perturbations. 

Pulina and Tacchella \cite{pulina2010abstraction} present CETAR: a Counter-Example Triggered Abstract Refinement 
verification approach for DNNs. Performance is not demonstrated on large NNs (only 20 nodes are used).

Paper \cite{pulina2011n} by the same authors describes and evaluates the tool NeVeR that verifies the safety of ANNs by 
encoding them as SMT-formula with linear inequalities. Furthermore, to improve scalability, the authors apply the 
abstraction refinement scheme presented in their earlier work.

Xiang, Tran and Johnson \cite{xiang2018output} present a verification method for multi-layer NNs and apply it to 
robotics. Their simulation-based method for the estimation of the output set of a NN is applicable to networks with 
monotone activation functions. The verification problem is formulated and solved as a chain of optimization problems for 
estimating the output-range.

Dutta et al. \cite{dutta2017output} also study the automatic estimation of the output-range for deep NNs. A key concept 
of theirs is that sets of possible inputs are compactly represented by convex polyhedral. They compute the guaranteed 
output range for DNNs by successive optimizations.

%%% NEW TEXT in Red Box
%\begin{tcolorbox}[colback=red!2!white, colframe=red!75!black]{ %Red Box
Baufreton et al. in 2010 \cite{baufreton2010multi} presented an analysis of safety standards and their implementation in 
certification strategies from different domains such as aeronautics, automation, automotive, nuclear, railway and space 
(performed in the context of the CG2E --- "Club des Grandes Entreprises de l'Embarqu\'e"). 
All the covered domains agree 
upon the articulation of a deterministic view of software and the system safety goals, including the probabilistic ones. 
The regulation regime and certification scheme is similar for aviation, nuclear and, to some extent, railway and space, 
but significantly different for automation and automotive.\\

Blanquart et al. in 2012 \cite{blanquart2012criticality} presented a comparative analysis across several industrial 
domains, of the fundamental notion of safety categories or levels (Safety Integrity Levels, Development Assurance 
Levels, etc.) underlying the safety framework enforced by safety standards, gathering experts from 6 industrial domains 
(automotive, aviation, industrial automation, nuclear, railway and space). They have shown that the various schemes are 
not fundamentally different, and could be seen as various instances of a single consistent scheme.\\

In the same 2012 Machrouh et al. presented an analysis of the impact of the Development Assurance Level (DAL) or Safety 
Integrity Level (SIL) on the system activities in various application domains represented in the CG2E and specially on 
the dependability, safety norms and standards working group. They analyzed the impact in each application domain, and 
identified and discussed the similarities and the dissimilarities in order to find the cross domain synergies.\\

Ledinot et al. in \cite{ledinot2012cross} compares the influence of Development Assurance Levels (DALs) of six different 
software development assurance standards for civil aviation, automotive, space, process automation, nuclear and railway. 
They observed significant cross-domain differences to minimize the risk of residual software development or verification 
errors. They found, that the discrepancies between the six standards in planning, in rules and standards, in structural 
coverage or verification independency etc. are not a matter of degree. Some major discrepancies are a matter of 
principles: definition of requirements vs. requirement of definitions, modulation of activities vs. modulation of means.\\

Seshia, Sadigh and Sastry \cite{seshia2016towards} analyzed the challenge of formally verifying systems that use 
artificial intelligence or machine learning. They identified five main challenges: environment modeling, formal 
specification, system modeling, computational engines, and correct-by-construction design. They are applying the 
developed theory to the design of human cyber-physical systems \cite{seshia2015formal} and learning-based cyberphysical 
systems, with a special focus on autonomous and semi-autonomous vehicles.
%} \end{tcolorbox} %End of the Red Box

%%% NEW TEXT in Red Box
%\begin{tcolorbox}[colback=red!2!white, colframe=red!75!black]{ %Red Box
In 2014, Ledinot et al. \cite{ledinot2014joint} discussed different approaches to combining formal methods (FM) and 
testing in the safety standards of the automotive, aeronautic, nuclear, process, railway and space industries. They 
concluded that Railway, Aeronautics, and to some extent Nuclear, are the three industrial domains where using formal 
methods, alone or jointly with testing, is effective in production software development. In case of joint use, three 
modes of combination may be considered, depending on whether one partitions, substitutes or intertwines the two 
verification means. Alternative and more direct means to address detection of unintended functions have been proposed 
formal methos (FM) 
verification of the specification, double independent specification, and enhanced exploratory testing in this paper. 
Then in 2016 \cite{ledinot2016perspectives} the authors propose a global rationale combining probabilistic evidence on 
hardware random failures and deterministic evidence on systematic causes of failures including software. They reject, 
for ultrahigh reliability software, a move towards more statistical assessment against less development assurance.\\

In the Best Paper of the ERTS$^2$ 2018 \cite{blanquart2018software}  the authors proposed a description of classical 
software safety analysis techniques, and discussed why software complexity increase has progressively made completeness 
of system functional safety requirements an important issue. They stress that extrapolating system or hardware analysis 
techniques such as Failure Modes and Effects Analysis (FMEA) to software is unlikely to provide meaningful results, 
considering that the underlying assumptions such as the fault model do not apply to software. However, techniques such 
as SEEA (Software Error Effect Analysis) may provide some support to robustness analysis. The proper development of 
pieces of software needs the generalization of techniques such as contract-based design with compositional verification, 
consistent safety invariants at all design levels, and a more control-oriented approach to safety.\\

Ruan, Huang and Kwiatkowska \cite{ruan2018reachability}  show how to obtain the safety verification problem, the output 
range analysis problem and a robustness measure by instantiating the reachability problem. They present a novel 
algorithm based on adaptive nested optimisation to solve the reachability problem. The technique has been implemented 
and evaluated on a range of deep neural networks (DNNs), demonstrating its efficiency, scalability and ability to handle 
a broader class of networks than state-of-the-art verification approaches.\\

%} \end{tcolorbox} %End of the Red Box

%%%%%%%%%%%%%%%%%%
% SPECIFICATION AND FUTURE SOFTWARE TOOLS
%%%%%%%%%%%%%%%%%%

\section{SPECIFICATION AND FUTURE SOFTWARE TOOLS}

The above set of research results indicate a strong convergence towards automatic and formally-based methods 
for 
verifying the input-output behavior of DL systems. But a serious problem appears to remain in balancing the 
guarantees 
of exhaustive search as in model checking with reasonable compute times. This situation is familiar to users 
of linear 
solvers and indeed several authors use linear equations and solvers to tackle DL safety problems. 

J. Taylor et al.'s paper \cite{taylor2016alignment} discusses in a very high-level way the problem of 
specifying the 
behavior of a machine-learning system for example through the objective function of its training phase. It 
covers an 
interesting set of research targets one of whom has specific meaning for specification of DL system behavior. 
Inductive 
ambiguity identification is defined as the goal of creating systems that can detect inputs for which their 
inference or 
classification would be highly under-determined by training data. 
Future safety-verification methods should 
address this problem that is akin to the need for attaching confidence levels to DL-system outputs. 

Foerster et al. \cite{foerster2016input} present a very innovative approach where the NNs come from a 
specific sub-family: without nonlinearities or input-dependent recurrent weights. For this family the linear 
representation of input-output behavior is not an approximation but an exact encoding. 
As a result verification can benefit from fast 
linear-algebra operations. The balance between this restricted family of NNs and their expressive power is 
illustrated on a very large NLP example. 
This approach could either become a breakthrough or a less significant approach for niche 
applications. 
But the general idea of a compact and efficiently-processed specification has been demonstrated.  

The white paper by Russel, Dewey and Tegmark \cite{russell2015research} reasserts, among many other things, 
that formal 
verification and security and absolute necessities for all AI systems. They propose that AI systems (among 
them DL 
systems) should allow the verification of their behavior, of their designs (in particular their specification)
, allow 
how to distinguish their software-hardware components, and also the modular verification of their parts.

%\begin{tcolorbox}[colback=red!2!white, colframe=red!75!black]
{ Cheng et al. \cite{cheng2018nn} presents the open-source toolbox nn-dependability-kit  to support data-
driven 
engineering of neural networks for safety-critical domains. They provide evidence of uncertainty reduction 
in key 
phases of the product life cycle, ranging from data collection, training \& validation, testing \& 
generalization, to operation.
  The application of Gaussian noise changed the result of classification, 
  where the confidence of being "end of no overtaking zone" has dropped from the originally 
  identified 100\% to 16.6\%.
  \\

Kulkarni et al. \cite{kulkarni2015picture} present Picture, a probabilistic programming language for scene 
understanding 
that allows researchers to express complex generative vision models, while automatically solving them using 
fast general-purpose inference machinery. 
Picture provides a stochastic scene language that can express generative models 
for arbitrary 2D/3D scenes, 
as well as a hierarchy of representation layers for comparing scene hypotheses with 
observed 
images by matching not simply pixels, but also more abstract features (e.g., contours, deep neural network 
activations). Such a language certainly improves programming productivity but its improvement of 
safety or verification remains to be seen. 
\\

A last recent line of research is the design of domain-specific programming languages (DSLs) 
that provide a white-box view of 
predefined NN libraries and frameworks. They allow users to write explicit and portable code for 
neural-net layers, their topology (data-dependencies) and allow the compiler writers to concentrate 
on optimizations and architecture models. The publications we cite here are only a few early examples 
of this research and we cannot be exhaustive about it at the time of writing this survey (2019Q1). 

A team from NVIDIA has presented its Diesel DSL \cite{Elango2018} specifically designed for producing 
efficient neural net implementations. The input Diesel program specifies a single-assignment set of arrays 
and data dependencies. It is compiled to a polyhedral intermediate representation allowing static 
data-size inference, layer (loop) fusions and tiling among other optimizations. 

In the context of the TVM software framework, another team has developed the Relay DSL 
\cite{Roesch2018} with even more ambitious language features to facilitate NN programming. 
It features a Python front-end for developers using that popular language, but more importantly: 
dependent types for tensor shapes, a TVM-integrated compiler, runtime optimizations and a module 
for automatic differentiation of programs. This last features is a the core of NN training procedures 
where a neural net's inference (execution) needs to be differentiated with respect to its error function. 
Training has previously been a mostly black-box operation from the point of view of source code. 
Training can now become explicit and source-code driven, thus lifting the effect of training to a 
province of programming language semantics. 

It should be hoped that program verification techniques 
will also evolve to make use of the precise semantics that can be attached to DSL operations. 

Among other research sub-directions that are completely open one can list: 
\begin{itemize}
  \item A DSL sub-language defining the distance function that is the basis for defining perturbations. 
  \item Tools that translate those DSLs into low-level specifications for given datasets, including tools to 
compare datasets, analyze them for their distance-function statistics etc. 
  \item UML class diagrams for representing datasets, others for replacing the DSLs in industrial 
applications. 
  \item Theorem-proving techniques are still far in the future because they require a clear 
logical specification of what a neural network's inference computes. 
\end{itemize}

%%%%%%%%%%%%%%%%%%
% Conclusion
%%%%%%%%%%%%%%%%%%

\section{Conclusion}

Safety of DL systems is a serious requirement for real-life systems and the research community is addressing 
this need 
with mathematically-sound but low-level methods that guarantee inference stability. 
But even when satisfactory and feasible, such a verification only guarantees that the original behavior of 
the given NN is unchanged from its training. 
Yet there are no verifiable guarantees that this is in itself correct and complete for lack of a 
specification. 

To turn DL system design into a broad 
industry, methods inspired by software engineering must be applied to complement current techniques. 

Our survey of the area has shown the acceleration of 
the line of work, the general agreement for its mathematical and low-level methods and their relative 
success as a first step in this direction.

%%%%%%%%%%%%%%%%%%
% ACKNOWLEDGMENT
%%%%%%%%%%%%%%%%%%

\section*{ACKNOWLEDGMENT}

Arvid Jakobsson is supported by a CIFRE industrial PhD contract between Huawei Technologies France and LIFO, 
Universit\'e d'Orl\'eans, funded in part by A.N.R.T. 
%The authors thank the reviewers for many relevant and/or insightful questions and corrections.

%\bibliographystyle{alpha}
%\balance
%\bibliography{Refs2019}

\begin{thebibliography}{10}
\providecommand{\url}[1]{#1}
\csname url@samestyle\endcsname
\providecommand{\newblock}{\relax}
\providecommand{\bibinfo}[2]{#2}
\providecommand{\BIBentrySTDinterwordspacing}{\spaceskip=0pt\relax}
\providecommand{\BIBentryALTinterwordstretchfactor}{4}
\providecommand{\BIBentryALTinterwordspacing}{\spaceskip=\fontdimen2\font plus
\BIBentryALTinterwordstretchfactor\fontdimen3\font minus
  \fontdimen4\font\relax}
\providecommand{\BIBforeignlanguage}[2]{{%
\expandafter\ifx\csname l@#1\endcsname\relax
\typeout{** WARNING: IEEEtran.bst: No hyphenation pattern has been}%
\typeout{** loaded for the language `#1'. Using the pattern for}%
\typeout{** the default language instead.}%
\else
\language=\csname l@#1\endcsname
\fi
#2}}
\providecommand{\BIBdecl}{\relax}
\BIBdecl

\bibitem{park2004content}
S.~B. Park, J.~W. Lee, and S.~K. Kim, ``Content-based image classification
  using a neural network,'' \emph{Pattern Recognition Letters}, vol.~25, no.~3,
  pp. 287--300, 2004.

\bibitem{kisavcanin2017deep}
B.~Kisa{\v{c}}anin, ``Deep learning for autonomous vehicles,'' in
  \emph{Multiple-Valued Logic (ISMVL), 2017 IEEE 47th International Symposium
  on}.\hskip 1em plus 0.5em minus 0.4em\relax IEEE, 2017, pp. 142--142.

\bibitem{Hains2018}
\BIBentryALTinterwordspacing
G.~Hains, A.~Jakobsson, and Y.~Khmelevsky, ``{Towards formal methods and
  software engineering for deep learning: Security, safety and productivity for
  {DL} systems development},'' in \emph{{2018 Annual IEEE International Systems
  Conference (SysCon)}}.\hskip 1em plus 0.5em minus 0.4em\relax Vancouver,
  Canada: {IEEE}, Apr. 2018. [Online]. Available:
  \url{https://hal.inria.fr/hal-01819035}
\BIBentrySTDinterwordspacing

\bibitem{carlini2017ground}
N.~Carlini, G.~Katz, C.~Barrett, and D.~L. Dill, ``Ground-truth adversarial
  examples,'' \emph{arXiv preprint arXiv:1709.10207}, 2017.

\bibitem{lu2017no}
J.~Lu, H.~Sibai, E.~Fabry, and D.~Forsyth, ``No need to worry about adversarial
  examples in object detection in autonomous vehicles,'' \emph{arXiv preprint
  arXiv:1707.03501}, 2017.

\bibitem{taylor2003verification}
B.~J. Taylor, M.~A. Darrah, and C.~D. Moats, ``Verification and validation of
  neural networks: a sampling of research in progress,'' in \emph{Intelligent
  Computing: Theory and Applications}, vol. 5103.\hskip 1em plus 0.5em minus
  0.4em\relax International Society for Optics and Photonics, 2003, pp. 8--17.

\bibitem{tian2018deeptest}
Y.~Tian, K.~Pei, S.~Jana, and B.~Ray, ``{DeepTest}: Automated testing of
  deep-neural-network-driven autonomous cars,'' in \emph{Proceedings of the
  40th International Conference on Software Engineering}.\hskip 1em plus 0.5em
  minus 0.4em\relax ACM, 2018, pp. 303--314.

\bibitem{leofante2016learning}
F.~Leofante, L.~Pulina, and A.~Tacchella, ``Learning with safety requirements:
  State of the art and open questions.'' in \emph{RCRA@ AI* IA}, 2016, pp.
  11--25.

\bibitem{wicker2018feature}
M.~Wicker, X.~Huang, and M.~Kwiatkowska, ``Feature-guided black-box safety
  testing of deep neural networks,'' in \emph{International Conference on Tools
  and Algorithms for the Construction and Analysis of Systems}.\hskip 1em plus
  0.5em minus 0.4em\relax Springer, 2018, pp. 408--426.

\bibitem{yerramalla2005stability}
S.~Yerramalla, \emph{Stability monitoring and analysis of online learning
  neural networks}.\hskip 1em plus 0.5em minus 0.4em\relax West Virginia
  University, 2005.

\bibitem{wu2018game}
M.~Wu, M.~Wicker, W.~Ruan, X.~Huang, and M.~Kwiatkowska, ``A game-based
  approximate verification of deep neural networks with provable guarantees,''
  \emph{arXiv preprint arXiv:1807.03571}, 2018.

\bibitem{dd2018ai2}
T.~Gehr, M.~Mirman, D.~Drachsler-Cohen, P.~Tsankov, S.~Chaudhuri, and
  M.~Vechev, ``Ai2: Safety and robustness certification of neural networks with
  abstract interpretation,'' in \emph{2018 IEEE Symposium on Security and
  Privacy (SP)}, 2018.

\bibitem{black2016sate}
P.~E. Black and A.~Ribeiro, ``{SATE V} {Ockham} sound analysis criteria,''
  \emph{National Institute of Standards and Technology (NIST). NIST IR 8113},
  2016.

\bibitem{georgakis2017synthesizing}
G.~Georgakis, A.~Mousavian, A.~C. Berg, and J.~Kosecka, ``Synthesizing training
  data for object detection in indoor scenes,'' \emph{arXiv preprint
  arXiv:1702.07836}, 2017.

\bibitem{georgakis2016multiview}
G.~Georgakis, M.~A. Reza, A.~Mousavian, P.-H. Le, and J.~Kosecka, ``Multiview
  {RGB-D} dataset for object instance detection,'' \emph{arXiv preprint
  arXiv:1609.07826}, 2016.

\bibitem{lai2014unsupervised}
K.~Lai, L.~Bo, and D.~Fox, ``Unsupervised feature learning for {3D} scene
  labeling,'' in \emph{Robotics and Automation (ICRA), 2014 IEEE International
  Conference on}.\hskip 1em plus 0.5em minus 0.4em\relax IEEE, 2014, pp.
  3050--3057.

\bibitem{hinterstoisser2017pre}
S.~Hinterstoisser, V.~Lepetit, P.~Wohlhart, and K.~Konolige, ``On pre-trained
  image features and synthetic images for deep learning,'' \emph{arXiv preprint
  arXiv:1710.10710}, 2017.

\bibitem{jang2017objective}
U.~Jang, X.~Wu, and S.~Jha, ``Objective metrics and gradient descent algorithms
  for adversarial examples in machine learning,'' in \emph{Proceedings of the
  33rd Annual Computer Security Applications Conference}.\hskip 1em plus 0.5em
  minus 0.4em\relax ACM, 2017, pp. 262--277.

\bibitem{leofante2018automated}
F.~Leofante, N.~Narodytska, L.~Pulina, and A.~Tacchella, ``Automated
  verification of neural networks: Advances, challenges and perspectives,''
  \emph{arXiv preprint arXiv:1805.09938}, 2018.

\bibitem{broderick2005adaptive}
R.~Broderick, ``Adaptive verification for an on-line learning neural-based
  flight control system,'' in \emph{Digital Avionics Systems Conference, 2005.
  DASC 2005. The 24th}, vol.~1.\hskip 1em plus 0.5em minus 0.4em\relax IEEE,
  2005, pp. 6--C.

\bibitem{van2017challenges}
P.~Van~Wesel and A.~E. Goodloe, ``Challenges in the {{Verification}} of
  {{Reinforcement Learning Algorithms}},'' NASA Langley Research Center;
  Hampton, VA, United States, Tech. Rep., Jun. 2017.

\bibitem{fuller2006stability}
E.~J. Fuller, S.~K. Yerramalla, and B.~Cukic, ``Stability properties of neural
  networks,'' in \emph{Methods and Procedures for the Verification and
  Validation of Artificial Neural Networks}.\hskip 1em plus 0.5em minus
  0.4em\relax Springer, 2006, pp. 97--108.

\bibitem{bunel2017piecewise}
R.~Bunel, I.~Turkaslan, P.~H. Torr, P.~Kohli, and M.~P. Kumar, ``Piecewise
  linear neural network verification: a comparative study,'' \emph{arXiv
  preprint arXiv:1711.00455}, 2017.

\bibitem{gopinath2017deepsafe}
D.~Gopinath, G.~Katz, C.~S. Pasareanu, and C.~Barrett, ``{DeepSafe}: A
  data-driven approach for checking adversarial robustness in neural
  networks,'' \emph{arXiv preprint arXiv:1710.00486}, 2017.

\bibitem{katz2017reluplex}
G.~Katz, C.~Barrett, D.~L. Dill, K.~Julian, and M.~J. Kochenderfer, ``Reluplex:
  An efficient {SMT} solver for verifying deep neural networks,'' in
  \emph{International Conference on Computer Aided Verification}.\hskip 1em
  plus 0.5em minus 0.4em\relax Springer, 2017, pp. 97--117.

\bibitem{cheng2017verification}
C.-H. Cheng, G.~N{\"u}hrenberg, and H.~Ruess, ``Verification of binarized
  neural networks,'' \emph{arXiv preprint arXiv:1710.03107}, 2017.

\bibitem{ehlers2017formal}
R.~Ehlers, ``Formal verification of piece-wise linear feed-forward neural
  networks,'' in \emph{International Symposium on Automated Technology for
  Verification and Analysis}.\hskip 1em plus 0.5em minus 0.4em\relax Springer,
  2017, pp. 269--286.

\bibitem{goodfellow2017challenge}
I.~Goodfellow and N.~Papernot, ``The challenge of verification and testing of
  machine learning,'' 2017.

\bibitem{huang2017safety}
X.~Huang, M.~Kwiatkowska, S.~Wang, and M.~Wu, ``Safety verification of deep
  neural networks,'' in \emph{International Conference on Computer Aided
  Verification}.\hskip 1em plus 0.5em minus 0.4em\relax Springer, 2017, pp.
  3--29.

\bibitem{katz2017towards}
G.~Katz, C.~Barrett, D.~L. Dill, K.~Julian, and M.~J. Kochenderfer, ``Towards
  proving the adversarial robustness of deep neural networks,'' \emph{arXiv
  preprint arXiv:1709.02802}, 2017.

\bibitem{narodytska2017verifying}
N.~Narodytska, S.~P. Kasiviswanathan, L.~Ryzhyk, M.~Sagiv, and T.~Walsh,
  ``Verifying properties of binarized deep neural networks,'' \emph{arXiv
  preprint arXiv:1709.06662}, 2017.

\bibitem{pulina2010abstraction}
L.~Pulina and A.~Tacchella, ``An abstraction-refinement approach to
  verification of artificial neural networks,'' in \emph{International
  Conference on Computer Aided Verification}.\hskip 1em plus 0.5em minus
  0.4em\relax Springer, 2010, pp. 243--257.

\bibitem{pulina2011n}
------, ``{NeVer}: a tool for artificial neural networks verification,''
  \emph{Annals of Mathematics and Artificial Intelligence}, vol.~62, no. 3-4,
  pp. 403--425, 2011.

\bibitem{xiang2018output}
W.~Xiang, H.-D. Tran, and T.~T. Johnson, ``Output reachable set estimation and
  verification for multilayer neural networks,'' \emph{IEEE transactions on
  neural networks and learning systems}, no.~99, pp. 1--7, 2018.

\bibitem{dutta2017output}
S.~Dutta, S.~Jha, S.~Sanakaranarayanan, and A.~Tiwari, ``Output range analysis
  for deep neural networks,'' \emph{arXiv preprint arXiv:1709.09130}, 2017.

\bibitem{baufreton2010multi}
P.~Baufreton, J.~Blanquart, J.~Boulanger, H.~Delseny, J.~Derrien, J.~Gassino,
  G.~Ladier, E.~Ledinot, M.~Leeman, P.~Qu{\'e}r{\'e} \emph{et~al.},
  ``Multi-domain comparison of safety standards,'' in \emph{Proceedings of the
  5th international conference on embedded real time software and systems
  (ERTS2), Toulouse, France}, 2010.

\bibitem{blanquart2012criticality}
J.-P. Blanquart, J.-M. Astruc, P.~Baufreton, J.-L. Boulanger, H.~Delseny,
  J.~Gassino, G.~Ladier, E.~Ledinot, M.~Leeman, J.~Machrouh \emph{et~al.},
  ``Criticality categories across safety standards in different domains,''
  \emph{ERTS-2012, Toulouse}, pp. 1--3, 2012.

\bibitem{ledinot2012cross}
E.~Ledinot, J.-M. Astruc, J.-P. Blanquart, P.~Baufreton, J.-L. Boulanger,
  H.~Delseny, J.~Gassino, G.~Ladier, M.~Leeman, J.~Machrouh \emph{et~al.}, ``A
  cross-domain comparison of software development assurance standards,''
  \emph{Proc. of ERTS2}, 2012.

\bibitem{seshia2016towards}
S.~A. Seshia, D.~Sadigh, and S.~S. Sastry, ``Towards verified artificial
  intelligence,'' \emph{arXiv preprint arXiv:1606.08514}, 2016.

\bibitem{seshia2015formal}
------, ``Formal methods for semi-autonomous driving,'' in \emph{Proceedings of
  the 52nd Annual Design Automation Conference}.\hskip 1em plus 0.5em minus
  0.4em\relax ACM, 2015, p. 148.

\bibitem{ledinot2014joint}
E.~Ledinot, J.-P. Blanquart, J.-M. Astruc, P.~Baufreton, J.-L. Boulanger,
  C.~Comar, H.~Delseny, J.~Gassino, M.~Leeman, P.~Qu{\'e}r{\'e} \emph{et~al.},
  ``Joint use of static and dynamic software verification techniques: a
  cross-domain view in safety critical system industries,'' in
  \emph{Proceedings of the 7th European Congress on Embedded Real Time Software
  and Systems (ERTS2 2014), Toulouse, France}, 2014, pp. 5--7.

\bibitem{ledinot2016perspectives}
E.~Ledinot, J.-P. Blanquart, J.~Gassino, B.~Ricque, P.~Baufreton, J.-L.
  Boulanger, J.-L. Camus, C.~Comar, H.~Delseny, and P.~Qu{\'e}r{\'e},
  ``Perspectives on probabilistic assessment of systems and software,'' in
  \emph{8th European Congress on Embedded Real Time Software and Systems (ERTS
  2016)}, 2016.

\bibitem{blanquart2018software}
J.-P. Blanquart, E.~Ledinot, J.~Gassino, P.~Baufreton, J.-L. Boulanger,
  S.~Brouste, J.~Camus, C.~Comar, P.~Qu{\'e}r{\'e}, and B.~Ricque, ``Software
  safety-a journey across domains and safety standards,'' in \emph{9th European
  Congress on Embedded Real Time Software and Systems (ERTS 2018)}, 2018.

\bibitem{ruan2018reachability}
W.~Ruan, X.~Huang, and M.~Kwiatkowska, ``Reachability analysis of deep neural
  networks with provable guarantees,'' \emph{arXiv preprint arXiv:1805.02242},
  2018.

\bibitem{taylor2016alignment}
J.~Taylor, E.~Yudkowsky, P.~LaVictoire, and A.~Critch, ``Alignment for advanced
  machine learning systems,'' \emph{Machine Intelligence Research Institute},
  2016.

\bibitem{foerster2016input}
J.~N. Foerster, J.~Gilmer, J.~Chorowski, J.~Sohl-Dickstein, and D.~Sussillo,
  ``Input switched affine networks: An {RNN} architecture designed for
  interpretability,'' \emph{arXiv preprint arXiv:1611.09434}, 2016.

\bibitem{russell2015research}
S.~Russell, D.~Dewey, and M.~Tegmark, ``Research priorities for robust and
  beneficial artificial intelligence,'' \emph{Ai Magazine}, vol.~36, no.~4, pp.
  105--114, 2015.

\bibitem{cheng2018nn}
C.-H. Cheng, C.-H. Huang, and G.~N{\"u}hrenberg, ``nn-dependability-kit:
  Engineering neural networks for safety-critical systems,'' \emph{arXiv
  preprint arXiv:1811.06746}, 2018.

\bibitem{kulkarni2015picture}
T.~D. Kulkarni, P.~Kohli, J.~B. Tenenbaum, and V.~Mansinghka, ``Picture: A
  probabilistic programming language for scene perception,'' in
  \emph{Proceedings of the IEEE conference on computer vision and pattern
  recognition}, 2015, pp. 4390--4399.

\bibitem{Elango2018}
V.~Elango, N.~Rubin, M.~Ravishankar, H.~Sandanagobalane, and V.~Grover,
  ``Diesel: {DSL} for linear algebra and neural net computations on {GPU}s,''
  in \emph{Proceedings of {MAPL'18}}.\hskip 1em plus 0.5em minus 0.4em\relax
  {ACM}, 2018.

\bibitem{Roesch2018}
J.~Roesch, S.~Lyubomirsky, L.~Weber, J.~Pollock, M.~Kirisame, T.~Chen, and
  Z.~Tatlock, ``Relay: A new {IR} for machine learning frameworks,'' in
  \emph{Proceedings of {MAPL'18}}.\hskip 1em plus 0.5em minus 0.4em\relax
  {ACM}, 2018.

\end{thebibliography}
\bibliographystyle{IEEEtran}


\end{document}